\documentclass[conference]{IEEEtran}
\IEEEoverridecommandlockouts
\usepackage{cite}
\usepackage{amsmath,amssymb,amsfonts}
\usepackage{graphicx}
\usepackage{textcomp}
\usepackage{listings}
\usepackage{xcolor}
\usepackage{url} 
\usepackage{authblk}
\usepackage{subcaption}
\usepackage{float}

\def\BibTeX{{\rm B\kern-.05em{\sc i\kern-.025em b}\kern-.08em
    T\kern-.1667em\lower.7ex\hbox{E}\kern-.125emX}}
     
\definecolor{codegreen}{rgb}{0,0.6,0}
\definecolor{codegray}{rgb}{0.5,0.5,0.5}
\definecolor{codepurple}{rgb}{0.58,0,0.82}
\definecolor{backcolour}{rgb}{1.,1.,1.}

\lstdefinestyle{mystyle}{
  backgroundcolor=\color{backcolour}, commentstyle=\color{codegreen},
  keywordstyle=\color{magenta},
  numberstyle=\tiny\color{codegray},
  stringstyle=\color{codepurple},
  basicstyle=\ttfamily\footnotesize,
  breakatwhitespace=false,         
  breaklines=true,                 
  captionpos=b,                    
  keepspaces=true,                 
  numbers=left,                    
  numbersep=-5pt,                  
  showspaces=false,                
  showstringspaces=false,
  showtabs=false,                  
  tabsize=2
}

\begin{document}

\title{When Structure is Silent: Opportunities for Algorithmic Dispatch in Linear Algebra\\

%{\footnotesize \textsuperscript{*}Note: Sub-titles are not captured for https://ieeexplore.ieee.org  andshould not be used}
\thanks{This material is based upon work supported by the U.S. National Science Foundation under award Nos CNS-2346520, PHY-2028125, RISE-2425761, DMS-2325184, OAC-2103804, and OSI-2029670, by the Defense Advanced Research Projects Agency (DARPA) under Agreement No. HR00112490488, by the Department of Energy, National Nuclear Security Administration under Award Number DE-NA0003965 and by the United States Air Force Research Laboratory under Cooperative Agreement Number FA8750-19-2-1000. Neither the United States Government nor any agency thereof, nor any of their employees, makes any warranty, express or implied, or assumes any legal liability or responsibility for the accuracy, completeness, or usefulness of any information, apparatus, product, or process disclosed, or represents that its use would not infringe privately owned rights. Reference herein to any specific commercial product, process, or service by trade name, trademark, manufacturer, or otherwise does not necessarily constitute or imply its endorsement, recommendation, or favoring by the United States Government or any agency thereof. The views and opinions of authors expressed herein do not necessarily state or reflect those of the United States Government or any agency thereof." The views and conclusions contained in this document are those of the authors and should not be interpreted as representing the official policies, either expressed or implied, of the United States Air Force or the U.S. Government.}
\thanks{Accepted manuscript of the paper published in the 2025 IEEE High Performance Extreme Computing Conference (HPEC). DOI: \protect\url{https://doi.org/10.1109/HPEC67600.2025.11196484}. \textcopyright{} 2025 IEEE. Personal use of this material is permitted. Permission from IEEE must be obtained for all other uses, in any current or future media, including reprinting/republishing this material for advertising or promotional purposes, creating new collective works, for resale or redistribution to servers or lists, or reuse of any copyrighted component of this work in other works.}
}

\author{Emmanuel Lujan}
\author{Alan Edelman}
\affil{Computer Science \& Artificial Intelligence Laboratory, \protect\\
Massachusetts Institute of Technology, 
\protect\\ Cambridge, Massachusetts, USA.}
\affil{
\textit{eljn@mit.edu, edelman@mit.edu}
}

\maketitle

\begin{abstract}
Algorithmic dispatch is essential for performance in linear‑algebra–intensive systems.
A persistent challenge lies in the treatment of \textit{structured} matrices.
Although such matrices are often described as “sparse,” the term \textit{structured} is more precise, as it highlights exploitable properties—such as bandedness or triangularity—whose algorithmic advantages extend beyond sparsity alone. When the dispatch strategy leaves these structures unrecognized, valuable opportunities for optimization are lost.
Recent advances in generative AI offer the promise of linking these silent structures to more effective algorithmic and architectural choices, supplying much of the missing connective tissue in computational linear algebra.
However, AI‑synthesized dispatch strategies also raise important questions about their theoretical soundness.
This work introduces analytical criteria—grounded in time‑complexity analysis—to determine when structure‑aware dispatch delivers tangible gains. We examine the overheads of structure detection and data‑format conversion, characterizing their impact on \textit{speedup} and \textit{slowdown}.
We illustrate these concepts through a case study on \textit{LU} factorization applied to banded matrices stored in a dense format, demonstrating results that align with theoretical bounds and reveal substantial gains in both performance and memory usage.
These analyses underscore the need for more intelligent dispatch strategies to recognize and exploit silent structures—an underused path to high‑performance linear algebra.
\end{abstract}

\begin{IEEEkeywords}
matrix structure, data format, algorithmic dispatch, LU, computational linear algebra.
\end{IEEEkeywords}

\section{Introduction}

\subsection{Challenges in Algorithmic Dispatch for Linear Algebra}
\label{challenges}

In linear algebra-intensive software systems, algorithmic choice can be critical. An optimal selection often involves navigating inherent trade-offs among accuracy, efficiency, and memory usage. This decision-making process extends beyond identifying a general algorithmic class—such as \texttt{LU} factorization—and demands careful consideration of specialized variants (e.g., the Unsymmetric MultiFrontal Method (\texttt{UMFPACK}) \cite{umfpack2004}, \texttt{KLU} \cite{klu2010}, or banded \texttt{LU} via \texttt{xGBTRF} \cite{LAPACK}), solver-specific parameters, data formats (e.g., dense, sparse, banded), and preconditioners \cite{Greenbaum1989-sv}. Additional layers of complexity arise from choices about parallelization strategies (e.g., block sizes, partitioning schemes), arithmetic precision (e.g., single, double, or mixed) \cite{Higham_Mary_2022}, and hardware backends (e.g., CPU vs. GPU) \cite{alomairy2024dynamic, dagger2}.

The resulting design space is high-dimensional and characterized by intricate interdependencies that are often under-documented and challenging to benchmark \cite{novikov2025, Fawzi2022-bu, petabricks1, SmartSolve2025}. Even seasoned practitioners may struggle to navigate this landscape effectively. For domain scientists—who need linear algebra to “just work” as part of a broader pipeline—these challenges become even more acute. The default strategy is often to rely on general-purpose routines or legacy configurations, not because they are optimal, but because they are available and stable \cite{harris2020}. Moreover,  if linear algebra routines are peripheral to the project, the incentive to tune them vanishes. Time, staffing, and budget constraints often make detailed performance analysis infeasible. As a result, many systems continue to operate with inefficient defaults—despite decades of advances in algorithmic theory and software. This disconnect between what could be achieved and what is routinely deployed underscores the pressing need for intelligent, automated algorithmic selection mechanisms that lower the barrier to high-performance computing.

\subsection{Silent Structures}

One particularly persistent challenge concerns structured matrices \cite{golub2013matrix, WACO2023, potapczynski2023cola, Stylianou2023}. 
While it is tempting to label matrices as “sparse” due to the predominance of zeros, we are uncomfortable with the gratuitous use of that term when \textit{structured} would be more apt. The crux of the matter is not the percentage of zero entries, but rather the algorithmic implications of the matrix’s structure. 
For example, applying a general-purpose sparse solver to a tridiagonal matrix is almost certainly inefficient—despite the matrix’s sparsity—because solvers tailored to tridiagonal structure offer superior performance \cite{golub2013matrix}. 

In other scenarios, if structured matrices are stored in dense formats, the system is likely to default to dense solvers, which again results in suboptimal efficiency.
Converting these matrices to an appropriate structured representation prior to execution can yield substantial acceleration gains—even when using general-purpose sparse solvers.
This point is underscored in an official MATLAB website publication: \textit{“In linear algebra, the path to speed often relies on taking advantage of matrix structure and there is one aspect of the structure of tridiagonal matrices that is important -- the fact they are very sparse. We haven't sped this up in (version) 24a but it's worth pointing out that if you can use sparse matrices for these problems then you really should. It's so much faster, even taking into account the speed-ups shown above. The memory benefits are huge too!”} \cite{croucher2024backslash}. 

Another notable example illustrating the substantial impact of structure-aware solver selection on computational performance comes from empirical results in \texttt{LinearSolve.jl} \cite{LinearSolve2024}, where benchmarks demonstrate up to a 10× speedup in \texttt{LU} factorization for Laplace-type matrices \footnote{\url{https://docs.sciml.ai/LinearSolve/stable/tutorials/accelerating_choices/}}
\cite{rackauckas2019confederated,DifferentialEquations.jl-2017}.

In practice, however, structural advantages are often underutilized. The result is a silent inefficiency: the structure is there, but the software does not "hear" it.

\subsection{Algorithmic Dispatch for Linear Systems}
\label{backslash}

\begin{lstlisting}[caption={Julia implementation of the backslash operator in LinearAlgebra.jl \cite{Bezanson2017}, showcasing algorithm dispatch based on the matrix's structural features and data format conversions. The function selects specialized solvers—such as triangular or diagonal back-substitution, LU, or QR factorization—by inspecting the input matrix's elements for properties like triangularity and squareness, and performing data format conversions when beneficial.}, label={code:backslash}]
    function (\)(A::AbstractMatrix, B::AbstractVecOrMat)
        require_one_based_indexing(A, B)
        m, n = size(A)
        if m == n
            if istril(A)
                if istriu(A)
                    return Diagonal(A) \ B
                else
                    return LowerTriangular(A) \ B
                end
            end
            if istriu(A)
                return UpperTriangular(A) \ B
            end
            return lu(A) \ B
        end
        return qr(A, ColumnNorm()) \ B
    end
\end{lstlisting}

To understand how to leverage these structures, we can look to the well‑established example of automated solver selection for linear systems. In both \texttt{Julia} \cite{Bezanson2017} and \texttt{MATLAB} \cite{MATLAB}, the backslash operator ($\backslash$) serves as the primary interface for solving these kinds of systems. Internally, the implementations perform structural analysis of the coefficient matrix—examining attributes such as triangularity, symmetry, and dimensionality—to apply advantageous data format conversions and select an appropriate solver. \texttt{Julia}'s backslash internal dispatch logic is illustrated in Source Code~\ref{code:backslash}, while a corresponding example of MATLAB's approach is presented by Croucher in \cite{croucher2024backslash}. In \texttt{Julia}’s implementation, the structural analysis is based on inspecting the matrix elements rather than relying only on its declared data type. For example, in the diagonal case, three conditions must hold: first, the matrix must be square, verified by comparing its dimensions (m == n); second, \texttt{istril} inspects the elements to confirm the matrix is lower triangular; and third, \texttt{istriu} similarly checks that it is upper triangular. When all conditions are satisfied, the matrix is converted to a Diagonal type and format—a memory‑efficient representation—and subsequently solved using a specialized diagonal solver. These systems exemplify how rule‑based heuristics can achieve high performance by selecting algorithms and data formats that exploit the matrix's inherent structure.

\subsection{AI‑Driven Algorithmic Dispatch \& Validation Challenges}

Recent advances in artificial intelligence (AI) are reshaping the landscape of algorithmic discovery. For example, Google DeepMind’s AlphaEvolve \cite{novikov2025,Fawzi2022-bu} uncovered a new matrix‑multiplication method that surpasses Strassen’s 1969 breakthrough—marking the first improvement in over half a century.
In a similar vein, the DARPA–MIT SmartSolve project \cite{SmartSolve2025} aims to leverage large language models and high‑level, high‑performance programming to automate the generation of selection heuristics. 
Such systems have the potential to supply much of the missing connective tissue in computational linear algebra—especially the heuristics that determine optimal algorithmic and architectural choices based on the input matrix structure.

However, the integration of AI‑generated heuristics into algorithmic dispatch introduces critical questions about validation and theoretical grounding. In particular, exploiting matrix structure entails overheads that must be amortized, including the costs of structural analysis, data‑format conversion, and, at the implementation level, the introduction of new software dependencies. What guarantees—empirical or analytical—should these heuristics provide to be considered reliable?

\subsection{Contributions and Organization}
\label{contributions}

This study addresses the question above by providing analytical formulations—based on time complexity analysis—to assess when structural exploitation leads to measurable gains. We pay special attention to structured matrices stored in dense formats and analyze the trade-offs involved in recognizing and converting them to optimized representations.

To underscore the applied relevance, we present a case study centered on \texttt{LU} factorization applied to banded matrices stored in a dense format, illustrating the substantial speedups achievable across a broader class of structurally sparse problems in dense linear algebra.

The following sections are organized as follows: Section~\ref{section:DSA} introduces the structure-aware dispatch strategy and defines the \texttt{speedup} and \texttt{slowdown} metrics. Sections~\ref{section:dsa_lu} and~\ref{empirical-lu} analyze the strategy in the context of banded \texttt{LU} factorization and present supporting empirical results. Conclusions are drawn in Section~\ref{section:conclusion}.

\section{Structure-Aware Dispatch}
\label{section:DSA}

This section introduces a structure-aware dispatch strategy designed to optimize performance for structured matrices. We formalize the approach and provide performance metrics to evaluate its benefits relative to a generic conterpart.

\subsection{Cost Functions}

A structure-aware approach—such as the one outlined in Subsection~\ref{backslash}— proceeds in three steps:
\begin{enumerate}
    \item Analyze the elements of the input matrix to identify its structure (e.g., banded, tridiagonal, symmetric).
    \item Convert the matrix to a suitable data format if the identified structure suggests a performance advantage.
    \item Select and apply an algorithm specialized for the structured representation.
\end{enumerate}

While the third step—employing a specialized algorithm—is generally more efficient than relying on a generic alternative, the first two steps introduce overhead. Specifically, the cost of structure detection and data format transformation must be carefully weighed against the anticipated performance benefits of specialized execution. 

To analyze the cost-benefit trade-offs of structure-aware dispatch, we define the following cost functions:
\begin{itemize}
\item $I(N)$: the number of operations required to identify the structure of an $N \times N$ matrix.
\item $C(N)$: the number of operations required to convert the matrix to a structure-specific data format (e.g., banded, tridiagonal).
\item $F(N)$: the number of operations required to apply a generic algorithm (e.g., dense LU) to the matrix.
\item $f(N)$: the number of operations required to apply a structure-specialized algorithm (e.g., banded LU) to the reformatted matrix.
\end{itemize}

If structure exploitation is advantageous, the total number of operations is given by:
\begin{equation}
T_{\text{structured}}^{\text{true}}(N) = I(N) + C(N) + f(N).
\label{eq:t-struct-true}
\end{equation}
If structure is identified but specialization is not beneficial, the total cost becomes:
\begin{equation}
T_{\text{structured}}^{\text{false}}(N) = I(N) + F(N).
\label{eq:t-struct-false}
\end{equation}
Finally, in the absence of any structure‑aware strategy—i.e., under a generic strategy—the total cost simplifies to:
\begin{equation}
T_{\text{generic}}(N) = F(N).
\label{eq:t-gen}
\end{equation}

\subsection{Trade-offs and Constraints: Speedup and Slowdown}

When structure exploitation is beneficial (Eq.~\ref{eq:t-struct-true}), the effectiveness of the structure-aware approach with respect to a generic approach (Eq.~\ref{eq:t-gen}) can be quantified by the ratio:
\begin{equation}
\frac{F(N)}{I(N) + C(N) + f(N)}.
\end{equation}
A ratio greater than one indicates a \texttt{speedup} relative to the generic algorithm, while a ratio less than one implies a \texttt{slowdown} due to insufficient gains from specialization. Hence, the effectiveness of the approach requires the following constraint:
\begin{equation}
\begin{aligned}
\texttt{Speedup constraint:}\quad & I(N)+C(N)+f(N) \\
                                & < F(N).
\end{aligned}
\label{eq:su-con}
\end{equation}
In contrast, when structure is detected but specialization is not applied (Eq.~\ref{eq:t-struct-false}), the approach incurs an overhead due to structure analysis with respect to the generic approach (Eq.~\ref{eq:t-gen}). The corresponding \texttt{slowdown} can be expressed as:
\begin{equation}
\frac{I(N) + F(N)}{F(N)},
\end{equation}
which measures the relative increase in computational cost arising from structure identification. Then, the overhead remains bounded if:
\begin{equation}
\texttt{Slowdown constraint: } I(N) \leq F(N).
\label{eq:sd-con}
\end{equation}
To establish concrete bounds on potential \texttt{speedup} or \texttt{slowdown}, one must analyze specific algorithmic cases—such as \texttt{LU}, \texttt{Cholesky}, or \texttt{QR} factorizations—where cost functions can be characterized more precisely in terms of asymptotic or empirical complexity.

\section{Structure-Aware LU for Banded Matrices}
\label{section:dsa_lu}

In this section, we derive theoretical bounds for the structure-aware approach applied to \texttt{LU} factorization of banded matrices. This allows us to compute the maximum \texttt{speedup} and worst-case \texttt{slowdown}, and to highlight potential optimization opportunities.

\subsection{Bounds}

In widely used linear algebra systems, \texttt{LU} factorization for dense matrices is typically performed using the \texttt{DGETRF} routine, where
\begin{equation}
F(N) \sim O(N^3).
\label{eq:D-N3}
\end{equation}
In contrast, banded \texttt{LU} factorization is typically performed using solvers such as \texttt{DGBTRF}, which scale as $O(N \cdot P \cdot Q)$, where $P$ and $Q$ denote the upper and lower bandwidths, respectively \cite{golub2013matrix}. This reveals a wide range of possible computational complexities:
\begin{equation}
f(N) \sim O(N) \text{ to } O(N^3),
\label{eq:S-N-N3}
\end{equation}
where the minimum number operation is based on $O(P \cdot Q) \sim O(1)$. Additionally, the cost of identifying whether a matrix exhibits a banded pattern, denoted by $I(N)$, and the cost of converting a dense matrix to banded representation, denoted by $C(N)$, are both typically 
\begin{equation}
I(N), C(N) \sim O(N^2).
\label{eq:IC-N2}
\end{equation}

\begin{figure}[htbp]
    \centerline{\includegraphics[width=0.45\textwidth]{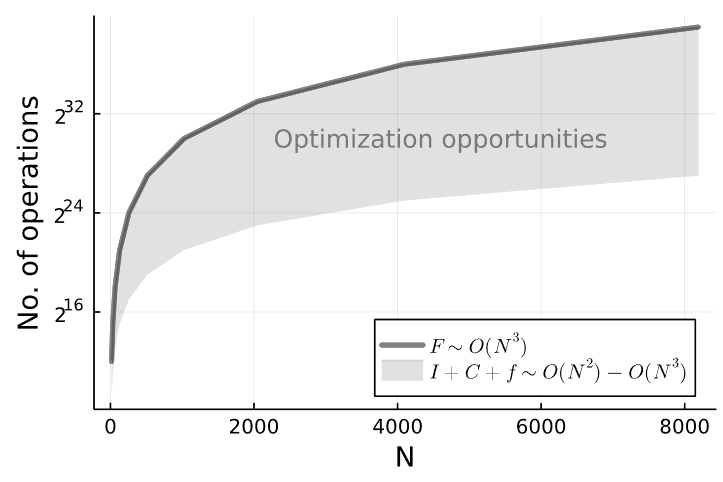}}
    \caption{Comparison of operation counts between generic (dark gray line) and structure-aware (light gray shaded region) approaches applied to \texttt{LU} and banded matrices. Structure-aware dispatch presents a wide range of optimization opportunities.}    
    \label{fig:operation_opportunities}
\end{figure}

Fig. \ref{fig:operation_opportunities} shows a comparison of operation counts between generic and structure-aware approaches applied to \texttt{LU}. The dark gray line shows the number of operations using the classical generic approach (Eq. \ref{eq:D-N3}) employed by commonly used libraries, where dense-formatted matrices are factorized using \texttt{DGETRF}. The light gray shaded region denotes the operational range of the structure-aware approach, in which the matrix pattern is first analyzed and then converted to an appropriate data format prior to applying a structured solver such as \texttt{DGBTRF}.
Based on Eq. \ref{eq:S-N-N3} and \ref{eq:IC-N2}, the number of operations of the structure-aware approach, $I(N)$+$C(N)$+$f(N)$, ranges from $O(N^2)$ to $O(N^3)$ .

\subsection{Maximum Speedup and Slowdown}

Based on the bounds introduced in the previous sub-section, the maximum theoretical \texttt{speedup} and \texttt{slowdown} of the structure-aware strategy, relative to the default dense approach, can be expressed as:

\begin{equation} \label{eq:speedupmax_lu}
    Speedup_{MAX} \sim
    \frac{O(N^3)}{O(N^2) + O(N^2) + O(N)} \sim O(N)
\end{equation} 

\begin{equation} \label{eq:slowdownmax_lu}
    Slowdown_{MAX} \sim \frac{O(N^2) + O(N^3)}{O(N^3)} \sim O(1)
\end{equation} 

\begin{figure}[htbp]
    \centerline{\includegraphics[width=0.45\textwidth]{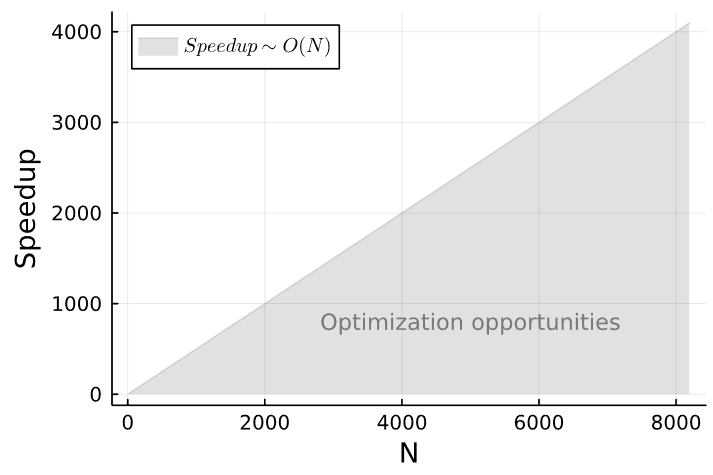}}
    \caption{Speedup opportunities of structure-aware over generic approaches in the context of \texttt{LU} factorization and banded matrices. Structure-aware dispatch presents a wide range of optimization opportunities.}
    \label{fig:speedup_opportunities}
\end{figure}

An $O(N)$ maximum \texttt{speedup} indicates that the performance can improve by up to a factor of $N$ compared to dense execution, while an $O(1)$ maximum \texttt{slowdown} implies that, in the worst case, the performance degradation remains constant and does not scale with the problem size $N$. The light gray region in Fig.~\ref{fig:speedup_opportunities} illustrates potential acceleration opportunities, with \texttt{LU} performance improvements scaling up to $O(N)$.

\section{Empirical results for LU}
\label{empirical-lu}

A simplified \texttt{Julia}-based implementation of the structure-aware strategy for \texttt{LU} factorization applied to banded matrices is presented in Source Code \ref{code:dsa_lu}. 

The function first inspects the pattern of the input dense matrix to determine whether it exhibits a banded structure. If such a structure is detected, the matrix is converted to an appropriate banded data format and a banded \texttt{LU} algorithm is applied; otherwise, a standard dense algorithm is used. This strategy enables adaptive algorithm selection based on the matrix's numerical structure rather than relaying on a default behavior.

%\newpage

\begin{lstlisting}[caption={Julia-based implementation demonstrating a structure-aware approach for data format conversion and algorithm selection applied to \texttt{LU} factorization of banded matrices.}, label={code:dsa_lu}]
    # Run LU on a dense-formatted matrix A
    function structure_aware_lu(A::Matrix)
        # Step 1: Identify matrix banded structure
        if banded_structure(A)
            # Step 2: Convert to banded format
            A_banded = BandedMatrix(A)
            # Step 3: Use banded algorithm: DGBTRF
            return lu(A_banded)
        else
            # Step 3: Use dense algorithm: DGETRF
            return lu(A)
        end
    end
\end{lstlisting}

\subsection{Empirical Speedup and Slowdown}

\begin{figure}[htbp]            
    \centerline{\includegraphics[width=0.45\textwidth]{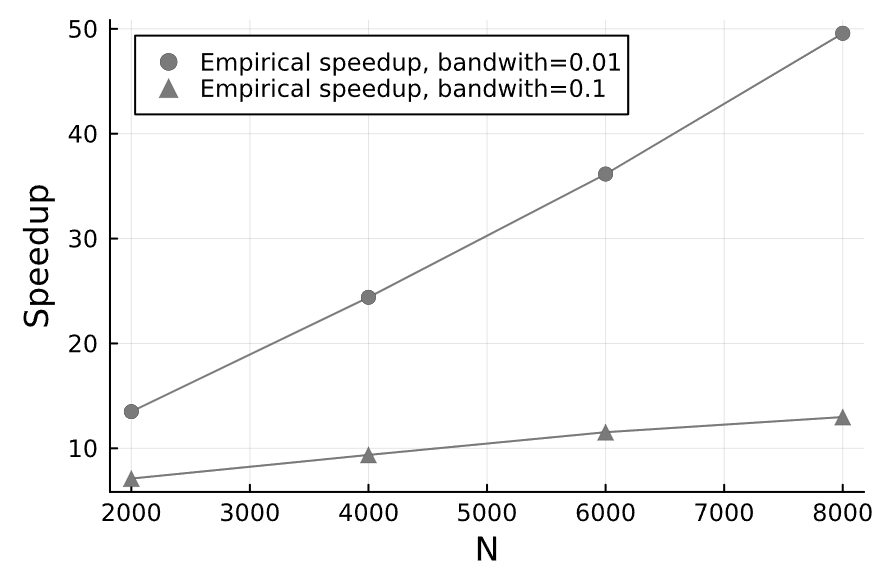}}
    \caption{ Empirical \texttt{speedup} achieved when structure is exploited and a specialized \texttt{LU} solver is used instead of the generic dense solver.  Results account for both the overhead of structure detection and conversion, and the performance gains from exploiting banded structure.}
    \label{fig:emp-speedup-lu}
\end{figure}

\begin{figure}[htbp]
    \centerline{\includegraphics[width=0.45\textwidth]{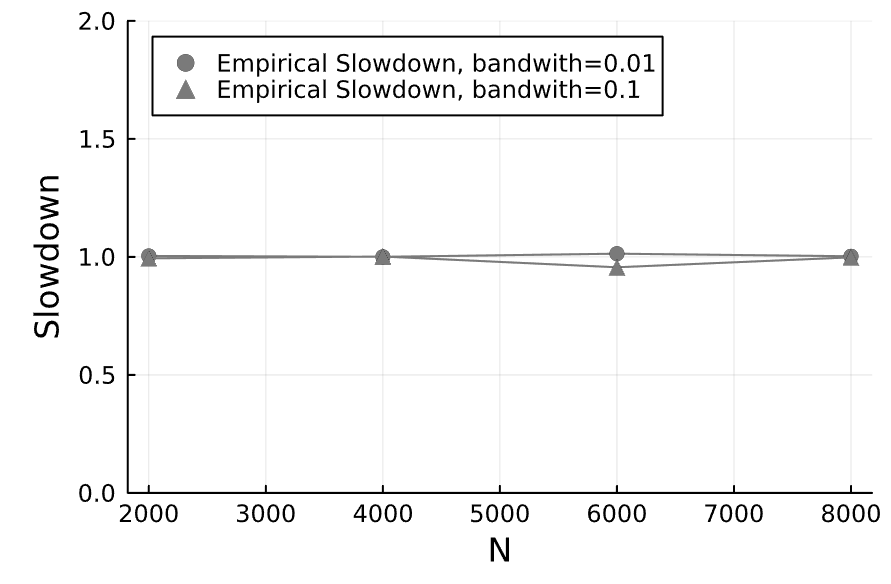}}
    \caption{Empirical \texttt{slowdown} observed when no exploitable structure is present and the dense \texttt{LU} solver is applied. Although structure detection introduces an $O(N^2)$ overhead, this cost is dominated by the $O(N^3)$ complexity of dense factorization, resulting in only a negligible, constant \texttt{slowdown}.}
    \label{fig:emp-slowdown-lu}
\end{figure}

Fig.~\ref{fig:emp-speedup-lu} and~\ref{fig:emp-slowdown-lu} show the empirical \texttt{speedup} and \texttt{slowdown} observed when executing the function defined in Source Code~\ref{code:dsa_lu} on Google Colab, using an Intel(R) Xeon(R) CPU @ 2.20GHz with 12.7~GB of RAM. We generate randomized banded matrices using the \texttt{BandedMatrices.jl} library and convert them to dense for benchmarking purposes. The size $N$ ranges from 2000 to 8000, and the bandwidths ($P+Q$) are set to 1\% and 10\% of each $N$. Execution times were measured using \texttt{BenchmarkTools.jl} \cite{BenchmarkTools.jl-2016}, and the results were summarized using median values. The figures reveal a linear \texttt{speedup} and a constant \texttt{slowdown} pattern, in agreement with the theoretical expectations given by Eq.~\ref{eq:speedupmax_lu} and~\ref{eq:slowdownmax_lu}, respectively. Moreover, the structure-aware strategy achieves accelerations ranging from 14X to 50X, with negligible overhead.

\subsection{Time and Memory Benchmarks}

This section reports time and memory benchmarks for the structure-aware function defined in Source Code~\ref{code:dsa_lu} and \texttt{Julia}’s standard \texttt{LU} routine, evaluated on different matrix configurations: a banded matrix stored in both banded and dense formats, and a dense matrix in dense format. 

\begin{figure}[htbp]
    \centering
    \begin{subfigure}[b]{0.45\textwidth}
        \centering
        \includegraphics[width=\textwidth]{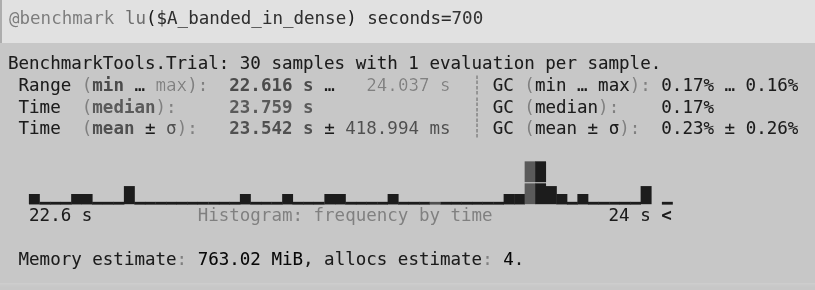}
        \caption{Standard \textit{LU} applied to a banded matrix in a dense format.}
        \label{fig:lu_banded_in_dense}
    \end{subfigure}
    \hfill
    \begin{subfigure}[b]{0.45\textwidth}
        \centering
        \includegraphics[width=\textwidth]{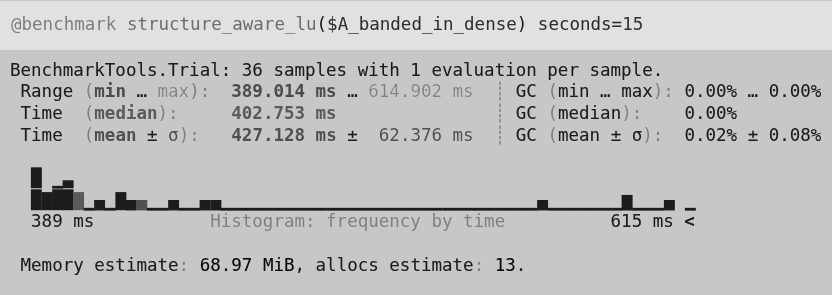}
        \caption{Structure-aware \textit{LU} applied to a banded matrix in a dense format.}
        \label{fig:palu_banded_in_dense}
    \end{subfigure}
    \caption{Julia-based benchmark of time and memory usage for \textit{LU} factorization on a banded matrix in dense format, comparing generic and structure‑aware dispatch.}
    \label{fig:lu_palu_banded_in_dense}
\end{figure}

The benchmarks in Fig.~\ref{fig:lu_palu_banded_in_dense} are based on a randomly generated $10^4 \times 10^4$ banded matrix with 100 sub-diagonals and 100 super-diagonals, subsequently converted to a dense format. Fig.~\ref{fig:lu_banded_in_dense} shows the performance of the standard \texttt{LU} function applied directly to the dense representation of the banded matrix, yielding a median execution time of 23.76 s and an estimated memory usage of 763.02 MiB. In contrast, Fig.~\ref{fig:palu_banded_in_dense} shows the result of applying the same operation using structure-aware dispatch, reducing the median execution time to 402.75 ms and memory usage to 68.97 MiB.
This results in a speedup of approximately 59×, in agreement with the theoretical time complexity analysis, along with a reduction in memory usage by a factor of 11.

\begin{figure}[htbp]
    \centering
    \begin{subfigure}[b]{0.45\textwidth}
        \centering
        \includegraphics[width=\textwidth]{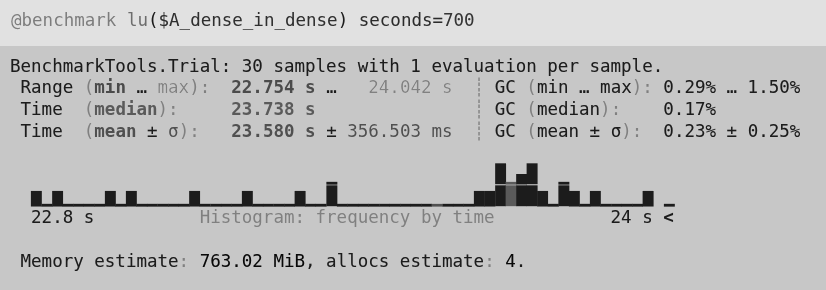}
        \caption{Standard \textit{LU} applied to a dense matrix in a dense format.}
        \label{fig:lu_dense_in_dense}
    \end{subfigure}
    \hfill
    \begin{subfigure}[b]{0.45\textwidth}
        \centering
        \includegraphics[width=\textwidth]{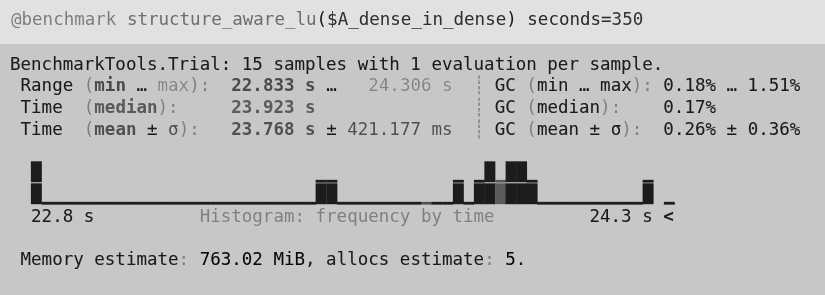}
        \caption{Structure-aware \textit{LU} applied to a dense matrix in a dense format.}
        \label{fig:palu_dense_in_dense}
    \end{subfigure}
    \caption{Julia-based benchmark of time and memory usage for \textit{LU} factorization on a dense matrix in dense format, comparing generic and structure‑aware dispatch.}
    \label{fig:lu_palu_dense_in_dense}
\end{figure}

The benchmarks in Fig.~\ref{fig:lu_palu_dense_in_dense} are also based on a randomly generated $10^4 \times 10^4$ dense matrix. Fig.~\ref{fig:lu_dense_in_dense} shows the performance of the standard \texttt{LU} function applied directly to the dense matrix, yielding a median execution time of 23.74 s and an estimated memory usage of 763.02 MiB. In contrast, Fig.~\ref{fig:palu_dense_in_dense} shows the result of applying the same operation using structure-aware dispatch, obtaining a median execution time of 23.92 ms and memory usage of 763.02 MiB. In this case, execution time and memory usage remain constant, aligning with the theoretical expectations.

\section{Conclusion}
\label{section:conclusion}

This study highlights the substantial yet often overlooked performance opportunities that emerge from exploiting structured matrices within algorithmic dispatch. As generative AI continues to advance, the emergence of new structure‑aware heuristics and algorithms will call for validation frameworks capable of assessing their soundness.

By formalizing the trade‑offs among structure detection, data‑format conversion, and specialized solver execution, we provide analytical criteria (Eqs. \ref{eq:su-con} and \ref{eq:sd-con}) to determine when a structure‑aware strategy offers measurable benefits. Applied to LU factorization of banded matrices, our analysis predicts a theoretical maximum speedup scaling as $O(N)$ with a worst‑case slowdown bounded by $O(1)$, results that are borne out by empirical benchmarks demonstrating significant acceleration and negligible overhead.

Overall, these analysis emphasize the critical role of recognizing and leveraging such silent structures as a pathway toward high‑performance linear algebra.

\section{Code Availability}

All code and data used in this study are publicly available at the following GitHub repository: \url{https://github.com/JuliaLabs/SmartSolve}. This repository contains the implementations, scripts, and instructions required to reproduce the results presented in this work.

\bibliographystyle{IEEEtran}
\bibliography{IEEEabrv,biblio}

\end{document}